\documentstyle[12pt,epsf]{article} 
\newcommand{\la}[1]{\label{#1}}
\newcommand{\nr}[1]{(\ref{#1})}
\newcommand{\lsi}{\raise0.3ex\hbox{$<$\kern-0.75em\raise-1.1ex\hbox{$\sim$}}}
\newcommand{\gsi}{\raise0.3ex\hbox{$>$\kern-0.75em\raise-1.1ex\hbox{$\sim$}}}
\newcommand{\lsim}{\mathop{\lsi}}
\newcommand{\gsim}{\mathop{\gsi}}
\renewcommand{\vec}[1]{{\bf #1}}
\newcommand{\be}{\begin{equation}}
\newcommand{\ee}{\end{equation}}
\newcommand{\stern}{^\ast}
\renewcommand{\(}{\left(}
\renewcommand{\)}{\right)}
\renewcommand{\[}{\left[}
\renewcommand{\]}{\right]}

\newcommand{\CA}{C_{A_{\scriptstyle \rm t}}}
\newcommand{\DA}{{\cal D}_{A_{\scriptstyle \rm t}}}

\begin{document}
 
\setlength{\baselineskip}{0.6cm}
\newcommand{\nn}{\nonumber}
\newcommand{\tr}{{\rm Tr\,}}
\newcommand{\fr}[2]{{\frac{#1}{#2}}}
\newcommand{\figysize}{16.0cm}
\newcommand{\figtopspace}{\vspace*{-1.5cm}}
\newcommand{\figbottomspace}{\vspace*{-5.0cm}}

\begin{titlepage}
\begin{flushright}
HD-THEP-97-32\\
hep-ph/9707489\\
October, 1997
\end{flushright}
\begin{centering}
\vfill
 
{\bf 
Plasmon properties in classical lattice gauge theory}
\vspace{1cm}

D. B\"odeker\footnote{bodeker@thphys.uni-heidelberg.de} and 
M. Laine\footnote{m.laine@thphys.uni-heidelberg.de} 

\vspace{1cm} {\em 
Institut f\"ur Theoretische Physik, Universit\"at Heidelberg, 
Philosophenweg 16, 
D-69120~Heidelberg, Germany}

\vspace{2cm}
 
{\bf Abstract}

\vspace{0.5cm}

In order to investigate the features of the classical
approximation at high temperatures for real time correlation
functions, the plasmon frequencies and damping rates were recently
computed numerically in the SU(2)+Higgs model and in the pure
SU(2) theory. We compare the lattice
results with leading order hard thermal loop resummed perturbation
theory. In the broken phase of the SU(2)+Higgs model,
we show that the lattice results can be
reproduced and that the lattices used are too coarse to observe some
important plasmon effects.  In the symmetric phase, the main
qualitative features of the lattice results can also be understood.
In the pure SU(2) theory, on the other hand, there are discrepancies
which might point to larger Landau and plasmon damping
effects than indicated by perturbation theory.
\end{centering}

\vspace{0.5cm}\noindent


\vspace{0.3cm}\noindent
 
\vfill \vfill
\noindent
 
\end{titlepage}
 


The dynamics of non-abelian gauge fields at finite temperature has
attracted some attention recently~\cite{grigoriev}--\cite{son}.  
A lot of work on this subject has been 
done in perturbation theory. However, some physical quantities
of interest are non-perturbative and the only known method of
calculating them on the lattice is the classical
approximation~\cite{grigoriev}.

One problem of the classical approximation is related to the
fact that the high momentum modes do not decouple from the dynamics of
the low momentum modes. In the naive perturbative expansion, momenta
of order $T$ lead to corrections in the Green's functions which are
proportional to $T^2$. For Green's functions with soft external
momenta $|\vec{p}|\sim g^2 T$, where $g$ is the gauge coupling, these
``corrections'' can be as large as or even larger than the tree level
Green's functions.  These large corrections have been named hard
thermal loops (HTL) and a consistent perturbative expansion requires
that they be resummed~\cite{pisarski}.

In the resummed perturbation theory the gauge field propagators
reflect two distinct collective phenomena. One is the plasma
oscillations which involve simultaneous oscillations of the low and
the high momentum degrees of freedom in the plasma.  The
characteristic frequency of these oscillations is proportional to
$gT$.  Secondly, there is an effect called
Landau damping which is related to an energy transfer from the low
momentum degrees of freedom to the high momentum ones.

In the classical approximation the physics of the high momentum
degrees of freedom is not correctly described. The hard thermal loops
correspond to classically UV-divergent contributions \cite{bodeker}. 
How the hard
thermal loops affect non-perturbative correlation functions of the low
momentum fields is an open problem.

The physical picture sketched above is based on 1-loop resummed
perturbation theory. It is therefore interesting to see whether it also
shows up in non-perturbative lattice studies. In a recent paper~\cite{tang} 
the correlation functions of two gauge invariant operators
in the SU(2)+Higgs model (in continuum notation),
\begin{eqnarray}
  H (t) &=& \frac{1}{\sqrt{V}}\int \!\! d^3 x \,\varphi^\dagger(t,\vec{x})
  \varphi(t,\vec{x}), \la{operH} \\
  W_i^a(t) &=& \frac{1}{\sqrt{V}}\int\!\! d^3 x \,
  i\tr\[\Phi^\dagger(t,\vec{x})
  D_i\Phi(t,\vec{x})\tau^a\], \la{operW}
\end{eqnarray}
were computed with classical lattice simulations.  Here $\varphi$
denotes the Higgs doublet, $\tau^a$ is a Pauli matrix, $\Phi$ is the
matrix $\Phi = (\widetilde\varphi\, \varphi)$ with
$\widetilde{\varphi} = i \tau^2 \varphi^*$, and $V$ denotes the space
volume.  In the broken phase the correlator of $W_i^a(t)$
is given by the gauge field propagator. It was claimed that the
plasmon frequency corresponding to this correlator is  
independent of the lattice
spacing $a$.
This contradicts the picture above since in the
classical theory the plasmon frequency is proportional to $T/a$
instead of $T^2$. We will demonstrate here that the qualitative
features of the correlators can be understood using perturbation
theory and that the claim of ref.~\cite{tang} concerning the 
broken phase $W_i^a$
plasmon frequency is not justified.

Ambj\o rn and Krasnitz \cite{ambjorn} have reported a measurement of
the gauge field correlator in the Coulomb gauge in the symmetric
phase.  We will see that many of the features observed can be
qualitatively understood in perturbation theory, but quantitative 
discrepancies remain. 

\subsubsection*{Real time correlators and the HTL effective action}

The plasmon properties of the 
classical Hamiltonian SU(2)+Higgs gauge theory could
be computed by solving the classical equations of motion
perturbatively with given initial conditions, 
and by then averaging over the initial conditions with 
the Boltzmann weight~\cite{grigoriev}. However, 
it appears that in perturbation theory it is simpler 
to perform the computation in the full quantum theory and
to take the classical limit $\hbar\to 0$ only in the end. 
We use this approach.

In the classical limit the operator ordering in a correlation 
function is irrelevant. Thus the quantum expression of which we are
going to take the
classical limit  can be anything but a commutator. 
This still leaves several possibilities, and we choose to consider
\begin{eqnarray}
  C_{\cal O}(t,\vec{p}) = \fr12 \int\! d^3 x e^{-i\vec{p}\vec{x}}
  \left\langle {\cal O}(t,\vec{x}) {\cal O}(0,\vec{0}) 
  + {\cal O}(0,\vec{0}) {\cal O}(t,\vec{x}) \right\rangle. \la{sym} 
\end{eqnarray}
In a perturbative
computation, one first computes the corresponding two-point Green's
function ${\cal D}_{\cal O}(i\omega_n,{\bf p})$ in Euclidean space for
the Matsubara frequencies $\omega_n$.  Then one performs an analytic
continuation which depends on the real time correlator in question.
The symmetric combination in eq.~\nr{sym} is given by
\begin{eqnarray}
        C_{\cal O}(t,{\bf p}) = 
        \hbar\int_{-\infty}^{\infty}\frac{d\omega}{2\pi}
        e^{-i\omega t}\left[1+2 n(\omega)\right]
        \mathop{\rm Im} {\cal D}_{\cal O}\left(
        i\omega_n\to \omega+i\epsilon,{\bf p}
        \right), \la{relation}
\end{eqnarray}
where
\begin{eqnarray}
  n(\omega)=\frac{1}{e^{\beta\hbar\omega}-1} \quad 
\end{eqnarray}
is the Bose distribution function. 

To evaluate the $\omega$-integral
in eq.\ (\ref{relation}) numerically, it is convenient to
rewrite it (for $t>0$) as an integral in the upper half of the complex
$\omega$-plane. In this way one avoids integrating along
the poles and discontinuities on the real $\omega$-axis. Denoting $\omega =
\omega_1 + i \omega_2$, one gets for $t>0$
in the classical limit
$n(\omega)\to T/(\hbar\omega)$, 
\begin{eqnarray}
        C_{\cal O}^{\rm classical}(t,{\bf p}) =  
        T \lim_{\hbar\to 0} {\cal D}_{\cal O}(0,{\bf p})+
        e^{\omega_2 t}\int_{-\infty}^{\infty}
        \frac{d \omega_1}{2\pi i}\frac{T}{\omega_1+i\omega_2}
        e^{-i\omega_1 t} \lim_{\hbar\to 0}
        {\cal D}_{\cal O}(\omega_1+i\omega_2,{\bf p}). \la{Ctp}
\end{eqnarray}
Note that this expression is independent of the imaginary part
$\omega_2>0$. The second term on the r.h.s.\ of eq.~\nr{Ctp} vanishes
for the equal time case $t=0$. A further virtue of eq.~\nr{Ctp} is
that one can take the limit ${\bf p}\to 0$ inside the integrand since
$\omega_2>0$ guarantees that there are no singularities on the
integration contour.  In the remainder of this letter we will discuss
only classical correlation functions\footnote{Except in eq.~\nr{AA}
and in the discussion thereafter.} and we will therefore omit the
superscript 'classical' from the l.h.s.\ of eq.~\nr{Ctp}.

We thus have to evaluate ${\cal D}_{\cal O}(\omega,{\bf p})$. As 
discussed above, a consistent perturbative expansion requires the
resummation of the hard thermal loops~\cite{pisarski}. 
In other words, the high momentum modes $p\gsim T$
(or $p\gsim a^{-1}$ in the classical theory) are integrated out, 
giving an effective theory for the low momentum modes $p\lsim gT$.
This amounts to
including the hard thermal loops in the tree level action. 
Denoting $P=(\omega_n,{\bf p})$, 
the quadratic part of the Euclidean
HTL effective theory in momentum space is 
\begin{eqnarray}
  {\cal L}_{\rm HTL}^{(2)} = \fr12 A_\mu^a A_\nu^a \left[ P^2
  \delta^{\mu\nu}-P^\mu P^\nu+\Pi_{\rm HTL}^{\mu\nu}(P) \right]
  +\varphi^\dagger \varphi \left[P^2 - \fr12 m_H^2 + \Sigma_{\rm
    HTL} \right], \la{htl}
\end{eqnarray}
where $m_H$ is the Higgs mass. Eq.~\nr{htl} gives 
directly the gauge and scalar field free propagators
${\cal D}_{\mu\nu}^{ab}(\omega, \vec{p})$, 
${\cal D}(\omega, \vec{p})$, to be used in 
the computation of ${\cal D}_{\cal O}(\omega,{\bf p})$.
Note that $\Sigma_{\rm HTL}$ is momentum independent.

Of the HTL self-energies, 
we will need especially the components $\Pi_{\rm HTL}^{ij}(P)$, 
$\Sigma_{\rm
HTL}$. The expressions are~\cite{pisarski,bodeker,arnold97}, 
after the replacement $\omega_n\to -i\omega$,
\begin{eqnarray}
  \Pi_{\rm HTL}^{ij}(P) & = & 
   -g^2 (2 N + N_{\rm s})\hbar   \int\!\! \frac{d^3q}{(2\pi)^3} \,
  \frac{dn(\omega_q)}{d\omega_q} \frac{v^iv^j\omega}{\omega- p^iv^i},
  \la{PiW} \\
   \Sigma_{\rm HTL} & = & \(6 \lambda + \frac94 g^2\) \hbar
  \int\!\! \frac{d^3q}{(2\pi)^3}\,\frac{n(\omega_q)}{\omega_q}, \la{PiH}
\end{eqnarray}
where, for the SU(2)+Higgs model, $N=2$, $N_{\rm s}=1$. Furthermore, 
\begin{eqnarray}
 v^i = \frac{\partial \omega_q }{\partial q_i} 
\end{eqnarray}
where $\omega_q$ is the tree level dispersion relation; for a space-time
with discretized spatial dimensions (a cubic lattice), 
\begin{equation} 
    \omega_q = \frac{2}{a} \sqrt{{\textstyle
    \sum_i}\sin^2\!  \left(a q_i/2 \right)}, \quad v^i = 
    \frac{\sin (a q^i)}{a\omega_q}.  
\end{equation} 
Eqs.~\nr{PiW}, \nr{PiH} hold actually for a generic
dispersion relation~\cite{arnold97}.  In ref.~\cite{bodeker}, scalar
electrodynamics was considered\footnote{The result for $\Pi_{\rm
    HTL}^{ij}$ given in~\cite{bodeker} has the wrong sign.} for which
case $g^2(2N+N_{\rm s})$ has to be replaced with $ 2e^2$ in
eq.~\nr{PiW}.

Note that in the theory of eq.~\nr{htl}, the gauge field $A_0$ is
included. In the classical lattice
simulations~\cite{tang,ambjorn}, in contrast, one puts $A_0=0$ and
imposes the Gauss constraint explicitly. 

The hard thermal loop self-energy simplifies 
in the limit ${\bf p}=0$, which is relevant 
for the operators in eqs.~\nr{operH}, \nr{operW}.
In that limit, the gauge field self-energy is
\begin{eqnarray}
  \Pi_{\rm HTL}^{00}=0,\quad \Pi_{\rm HTL}^{0i}=0,\quad \Pi_{\rm
    HTL}^{ij}=\delta^{ij} \omega_W^2,
\end{eqnarray}
where $\omega_W$ is the plasmon frequency, 
\begin{eqnarray}
   \omega_W^2=-\fr13 g^2 (2N+N_{\rm s})\hbar 
   \int\! \frac{d^3q}{(2\pi)^3} 
   \frac{dn(\omega_q)}{d\omega_q}|{\bf v}|^2. \la{oW2}
\end{eqnarray}
Let us also introduce the ``scalar plasmon frequency''
\begin{eqnarray}
\omega_H^2 = -\fr12 m_H^2 + \Sigma_{\rm HTL}. \la{oH2}
\end{eqnarray}
In the continuum limit of the full quantum
theory, for which $|{\bf v}|^2=1$, one has
\begin{eqnarray}
   \omega_W^2 & = &  g^2(2 N+N_{\rm s})\frac{T^2}{18}
   \frac{1}{\hbar},\la{omegaWT} \\
   \omega_H^2 & = &  -\fr12 m_H^2+
   \left(6\lambda + \fr94g^2\right)\frac{T^2}{12}
   \frac{1}{\hbar}, \la{omegaHT}
\end{eqnarray}
whereas in the classical limit $n(\omega)\to T/(\hbar\omega)$
on a cubic lattice, one gets
\begin{eqnarray}
  \omega_W^2 & = & g^2 (2 N+N_{\rm s}) \left(\fr32\frac{\Sigma}{\pi}-1
  \right) \frac{T}{12a},\la{omegaWa} \\ 
  \omega_H^2 & = & -\fr12 m_H^2 +
  \left(6\lambda + \fr94g^2\right) \frac{\Sigma}{4\pi}\frac{T}{a}. \la{omegaHa}
\end{eqnarray}
Here
\begin{eqnarray}
\Sigma =  \frac{1}{\pi^2}\int_{-\pi/2}^{\pi/2}\!\! d^3x
\frac{1}{\sum_i{\sin}^2x_i}= 3.175911535625 \la{Sigma}
\end{eqnarray}
is a constant 
which can be expressed in 
terms of the complete elliptic integral 
of the first kind~\cite{fkrs}.

The basic problem of the classical real time simulations can now be
expressed as follows. For the scalar field, one has a mass parameter
in the classical SU(2)+Higgs theory which does not break gauge
invariance. It can be tuned such that the high momentum modes produce
the correct quantum expression in eq.~\nr{omegaHT}, and that the
lattice spacing dependence in eq.~\nr{omegaHa} is canceled. For the
gauge fields, in contrast, the local classical SU(2)+Higgs theory does
not allow a mass term and hence the divergent classical expression in
eq.~\nr{omegaWa} cannot be arranged to coincide with the quantum
expression in eq.~\nr{omegaWT}.  This divergence should also show up
in the classical simulations. This problem is specific for time
dependent correlation functions: in the static case only the
$\omega_n=0$ sector requires a resummation and 
$\Pi_{\rm HTL}^{ij}(0,\vec{p})=0$.

Finally, let us fix some notation.
The continuum field $\varphi$ 
in eqs.~\nr{operH}, \nr{operW} is related
to the dimensionless
lattice field $\bar{\varphi}$ in~\cite{tang} by
\begin{eqnarray}
\bar\varphi^\dagger\bar\varphi=
\frac{a}{T\beta_G}\varphi^\dagger\varphi, \quad \beta_G =
\frac{4}{ag^2T}.
\end{eqnarray}
The operators measured on the lattice are then
\begin{eqnarray}
\bar H & = & \frac{1}{\sqrt{N^3}}\sum_{\bf x}
\bar\varphi^\dagger({\bf x})\bar\varphi({\bf x}), \\
\bar{W}^a_i & = &  
\frac{1}{\sqrt{N^3}}\sum_{\bf x}
i\tr \left[\bar\Phi^\dagger(\vec{x}) U_i(\vec{x})
\bar\Phi(\vec{x}+\vec{e}_i a)\tau^a\right],
\end{eqnarray}
where $U_i(\vec{x})$  
is the link operator. 
The correlators measured with these operators 
are denoted by $C_{\bar H}(t)$, 
$\fr13\delta^{ab}\delta_{ij}C_{\bar W}(t)$, and
they are related in the continuum limit $a\to 0$ 
to the correlators 
$C_H(t), C^{ab}_{W,ij}(t)$
of the continuum operators in eqs.~\nr{operH}, \nr{operW}
through
$C_H(t)=a \beta_G^2 T^2 C_{\bar H}(t)$,
$C^{ab}_{W,ij}(t)=\fr13
\delta^{ab}\delta_{ij} a^{-1} \beta_G^2 T^2 C_{\bar W}(t)$.
The factor $3$ in the definition of $C_{\bar W}(t)$ corresponds to 
a sum over the different isospin components.

\subsubsection*{The broken phase}

Let us first consider the broken phase. The real time correlators of
the operators $\bar H(t)$, $\bar W^a_i(t)$ have been determined in the
classical approximation in the broken phase of the SU(2)+Higgs model
by Tang and Smit~\cite{tang}.

We parameterize the scalar field in the broken phase as
\begin{eqnarray}
        \Phi=\frac1{\sqrt{2}}\[v(T) + \phi_0+ i\tau^a\phi_a\].
\end{eqnarray}
Then, in continuum notation, the lowest order terms in the gauge invariant
operators of eqs.~\nr{operH}, \nr{operW} become
\begin{eqnarray}
   H(t) & = & {\rm const.} + \frac{1}{\sqrt{V}}
   \int \!\!d^3x\, v(T)\phi_0(t,{\bf x}) , \\
   W^a_i (t) & = & \frac{1}{\sqrt{V}}
   \int \!\!d^3x\, v(T) \Bigl[ m_W(T) A^a_i(t,{\bf x}) 
   - \partial_i \phi_a(t,{\bf x}) \Bigr], \la{lWa}
\end{eqnarray}
where $m_W(T)=gv(T)/2$.  Due to the integration $\int\! d^3x$, these
operators correspond to zero spatial momentum, ${\bf p}=0$, so that
the second term in eq.~\nr{lWa} does not contribute.  It follows that
the leading terms in the correlation functions $C_H(t)$ 
and $C^{ab}_{W,ij}(t)$
are given by the propagators of the elementary fields
$\phi_0$ and $A^a_i$.

To compute the required propagators, one can set ${\bf p}=0$ in the time
dependent part of eq.~\nr{Ctp}, so that eqs.~\nr{oW2}, \nr{oH2} can be
used.  At zero spatial momentum the full Euclidean (Matsubara)
propagators for the fields $A^a_i$, $\phi_0$ are of the form
\begin{eqnarray}
   {\cal D}_{ij}^{ab}(i\omega_n, \vec{0}) = 
   \frac{\delta^{ab}\delta_{ij}}
   {\omega_n^2+(\omega_W^{\rm b})^{2}+  \mbox{$\stern\Pi(i\omega_n)$}},\quad
   {\cal D}(i\omega_n, \vec{0}) = 
   \frac{1}{\omega_n^2+(\omega_H^{\rm b})^{2}+ 
   \mbox{$\stern\Sigma(i\omega_n)$}},
\end{eqnarray}
where
$\stern\Pi$, $\stern\Sigma$ denote the parts of the self-energies
which are generated radiatively within the HTL effective theory. 
The tree-level terms are
\begin{eqnarray}
   (\omega_W^{\rm b})^2 & = & 
   m_W^2(T)+0.05379 \beta_G (g^2T)^2, \la{wbW} \\
   (\omega_H^{\rm b})^2 & = & m^2_H(T), \la{wbH}
\end{eqnarray}
where $m_W(T)$, $m_H(T)$ are the mass parameters generated by the
Higgs mechanism in the static theory: $m_W(T)=gv(T)/2$,
$m_H^2(T)=\omega_H^2+3 \lambda v^2(T)$.  Here the scalar mass
parameter was tuned to its correct value by using the counterterm of
the static classical theory and the fact that $\Sigma_{\rm HTL}$ is
momentum independent. For $(\omega_W^{\rm b})^2$, in contrast, the
$\beta_G$-dependent part comes from eq.~\nr{omegaWa} and cannot be
removed.  Parametrically, $(\omega^{\rm b})^{2}\sim g^2T^2$ and
$\stern\Pi(\omega_W^{\rm b})$, 
$\stern\Sigma(\omega_H^{\rm b})\sim g^3T^2$, so
that the dominant contributions to the $H$ and $W^a_i$ plasmon
frequencies (appearing as $C(t,{\bf p}) \sim C_0 \exp({-\Gamma t})
\cos(\omega_{\rm pl} t + \delta)$ in the correlator) are just
$\omega_H^{\rm b}$ and $\omega^{\rm b}_W$ according to eq.~\nr{Ctp}.
The damping rates $\Gamma$ are related to 
$\stern\Pi(\omega_W^{\rm b})$
and $\stern\Sigma(\omega_H^{\rm b})$.

\begin{figure}[tb]
 
\vspace*{-1.0cm}
 
\hspace{1cm}
\epsfysize=18cm
\centerline{\epsffile{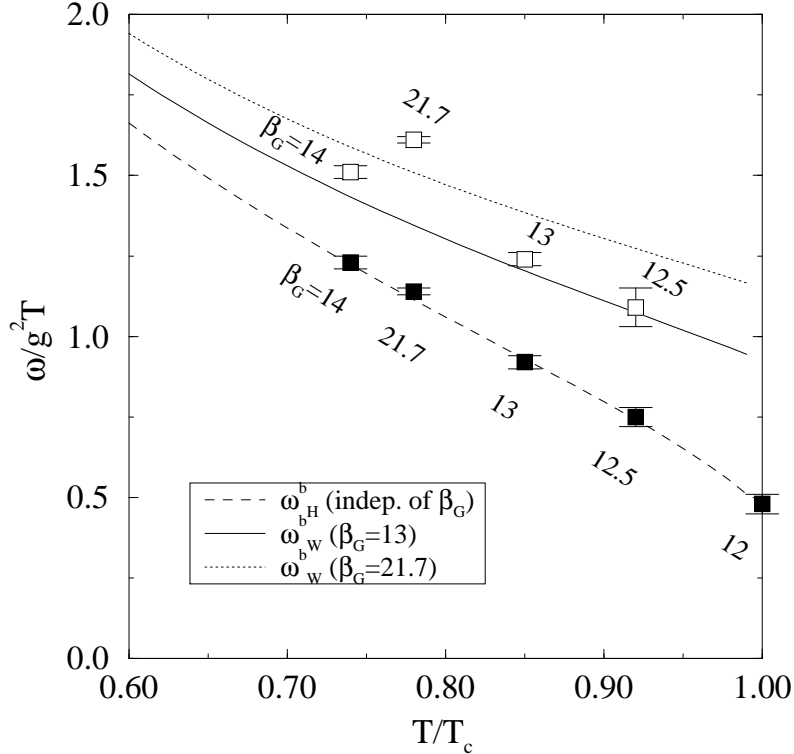}}
 
\vspace*{-6cm}
 
\caption[a]{
A comparison of the leading order perturbative
plasmon frequencies with those determined on the 
lattice in~\cite{tang}. The continuous curves are the
perturbative values from eqs.~\nr{wbW}, \nr{wbH}, and the squares 
are the data points from Table~5 in~\cite{tang}.
Open symbols correspond to $\omega_W^{\rm b}$ and
filled to $\omega_H^{\rm b}$. The corresponding
$\beta_G$ values are given next to the symbols.}
\la{omega}
\end{figure}

In Fig.~\ref{omega} we compare the leading order plasmon frequencies
in eqs.~\nr{wbW}, \nr{wbH} with the lattice results of
ref.~\cite{tang}. The value of $v(T)$ has been determined from the
1-loop effective potential.  The zero-temperature parameters used
in~\cite{tang} correspond to $m_H$ = 80 GeV.  It is seen that the
lattice results are remarkably close to the leading order
perturbative results. In particular, we conclude that the gauge field
plasmon frequency $\omega_W^{\rm b}$ diverges in the continuum limit
according to eq.~\nr{wbW}, while the scalar plasmon frequency
$\omega_H^{\rm b}$ remains finite and equals the static screening mass
at leading order. One sees that the lattice is so coarse that it is
difficult to notice the divergence of $\omega_W^{\rm b}$ since this is
shadowed by the finite $m_W(T)$.  Thus one is in a sense not close
enough to the continuum limit.  It should also be noted that the
amplitude of $C^{ab}_{W,ij}(t)$ dies out as $1/(\omega_W^{\rm b})^{2}$
in the continuum limit.

The damping rates of the gauge and scalar fields are parametrically of order 
$g^2T$ and therefore, 
in contrast to the plasmon frequencies,
they are classical. This has been demonstrated  
explicitly in a scalar field theory~\cite{aarts,jakovac}.
A full computation of the damping rates 
in the broken phase of the SU(2)+Higgs theory
is missing at the moment. However, the order of magnitude can 
apparently be understood~\cite{tang} using the known symmetric phase 
gauge and Higgs elementary field damping rates~\cite{braaten,biro}.

\subsubsection*{The symmetric phase}

Let us now turn to the symmetric phase. 
The correlators of the composite operators in eqs.~\nr{operH}, 
\nr{operW} have been determined in the symmetric phase
of the SU(2)+Higgs model by Tang and Smit~\cite{tang}.
In addition, the gauge field correlator of the pure SU(2) 
model has been measured in the Coulomb gauge by 
Ambj{\o}rn and Krasnitz~\cite{ambjorn}.

Consider first the composite operator correlators measured
in~\cite{tang}. In the symmetric phase, 
the composite operator character of $H(t)$ and $W^a_i(t)$
manifests itself more clearly than in the broken phase. 
For instance, the leading term in $\bar W^a_i$ is 
\begin{eqnarray}
  \bar W^a_i = \frac{1}{\sqrt{N^3}}\sum_\vec{x} \left[
  \bar\phi_0(\vec{x} + 
  \vec{e}_i a)\bar\phi_a(\vec{x})-\bar\phi_0(\vec{x})\bar\phi_a(\vec{x} + 
  \vec{e}_i a)-\epsilon^{abc}
  \bar\phi_b(\vec{x})\bar\phi_c(\vec{x} + \vec{e}_i a)\right],
\end{eqnarray}
which does not contain the gauge field $A^a_i$ at all.

\begin{figure}[tb]
 
\vspace*{-4.0cm}
 
\hspace{1cm}
\epsfysize=30cm
\centerline{\epsffile{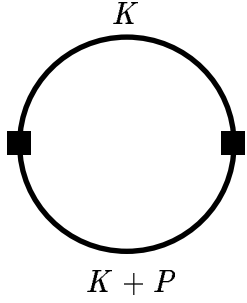}}
 
\vspace*{-22cm}
 
\caption[a]{The lowest order diagram 
contributing  to $C_H(t)$ and $C^{ab}_{W,ij}(t)$ in the symmetric phase.
The solid lines denote scalar propagators.}
\label{diagram}
\end{figure}

Due to the fact that no gauge fields are involved, the leading terms in
$C_H(t)$ and $C^{ab}_{W,ij}(t)$ can be easily computed: both are given
by diagrams of the type depicted in Fig.~\ref{diagram}. To evaluate
them one has to start with a Matsubara external momentum $p_0 =
i\omega_n$. Then the sum over the loop frequencies is written in terms
of an integral in the complex $k_0$ plane (see, e.g.,~\cite{kapusta}).
Only then can one continue to arbitrary complex values of $p_0$.  For
the operator ${\cal O}(t,\vec{x}) =
\varphi^\dagger\varphi(t,\vec{x})$, the diagram in Fig.~\ref{diagram}
finally gives
\begin{eqnarray}
  & & {\rm Im} {\cal D}_{\varphi^\dagger\varphi}(p_0 + i\epsilon,\vec{p}) 
  = \nn \\ 
  & & \quad\quad
  4 \int\! \frac{d^4k}{(2\pi)^4}
  \[n(k_0) - n(k_0+p_0) \]
  {\rm Im} {\cal D}(k_0 +i\epsilon, \vec{k})\, 
  {\rm Im} {\cal D}(k_0 + p_0 + i\epsilon, \vec{k} + \vec{p}). 
\hspace*{1.0cm}
\label{imDphi2}
\end{eqnarray}
This expression\footnote{Eq.~\nr{imDphi2} can be written 
in other forms by a change of integration variables, but
then one may get a wrong result for the 
free case, if $\epsilon$ is put to zero inside the integral
and eq.~(\ref{help1}) is used.} is equivalent to 
a corresponding one derived in ref.~\cite{jeon}. 
The correlator $C_H(t)=\fr12\langle H(t)H(0)+H(0)H(t)\rangle$
for the operator $H(t)$ of eq.~\nr{operH}
is given by $C_H(t)=C_{\varphi^\dagger\varphi}(t,\vec{p}=0)$, 
where $C_{\varphi^\dagger\varphi}(t,\vec{p})$ is obtained from 
eq.~\nr{relation}.

To get the leading contribution in $C_H(t)$, one uses the 
free propagators for which
\begin{eqnarray}
         {\rm Im} {\cal D}_{\rm free}(k_0 +i\epsilon, \vec{k}) = 
         \frac{\pi}{2\sqrt{\vec{k}^2 + \omega_H^2}}
        \[\delta\(k_0 - \sqrt{\vec{k}^2 + \omega_H^2}\) -
        \delta\(k_0 + \sqrt{\vec{k}^2 + \omega_H^2}\)\]. \label{help1}
\end{eqnarray}
Then the integrals over $k_0$ and $p_0$ can be performed
in eqs.~\nr{imDphi2}, \nr{relation}. 
In the limit $ \vec{p}\to 0$, one obtains
\begin{eqnarray}
  \label{chfree}
  C_H(t) = T^2 \int \frac{d^3k}{(2\pi)^3} \frac{1}{(\vec{k}^2 +\omega_H^2)^2}
  \left[1 + \cos\left(2\sqrt{\vec{k}^2 +\omega_H^2}\, t\right)\right].
  \la{loCHt}
\end{eqnarray}
On the lattice (in the continuum limit) this corresponds to 
\begin{eqnarray}
  C_{\bar H}(t)=
  \frac{1}{4 \pi^2\beta_G^2} 
  \int_0^{\infty} dx \frac{x^2}{(x^2+z^2)^2}\left[1+ \cos(4 \bar
  t\sqrt{x^2+z^2})\right], \la{symch}
\end{eqnarray}
where $z=a\omega_H/2$ and $\bar t=t/a$.  

The vector correlator 
$ C^{ab}_{W,ij}(t)$ has an additional factor of $k_ik_j$
in the integrand compared with eq.~\nr{loCHt}. 
Therefore the continuum limit of  $ C^{ab}_{W,ij}(t=0)$ does
not exist. On the lattice one obtains
\begin{eqnarray}
  C_{\bar W}(t) = \frac{2}{\pi^3\beta_G^2}
  \int_0^{\pi/2}\!d^3 x
  \frac{\sum_i\sin^2 2x_i}{[\sum_i\sin^2 x_i+z^2]^2}\left[1+
  \cos\left(4 \bar
  t \sqrt{{\textstyle \sum_i}\sin^2 x_i + z^2} \right)
  \right]. \la{symcw}
\end{eqnarray}

\begin{figure}[tb]
 
\vspace*{-1.0cm}
 
\hspace{1cm}
\epsfysize=18cm
\centerline{\epsffile{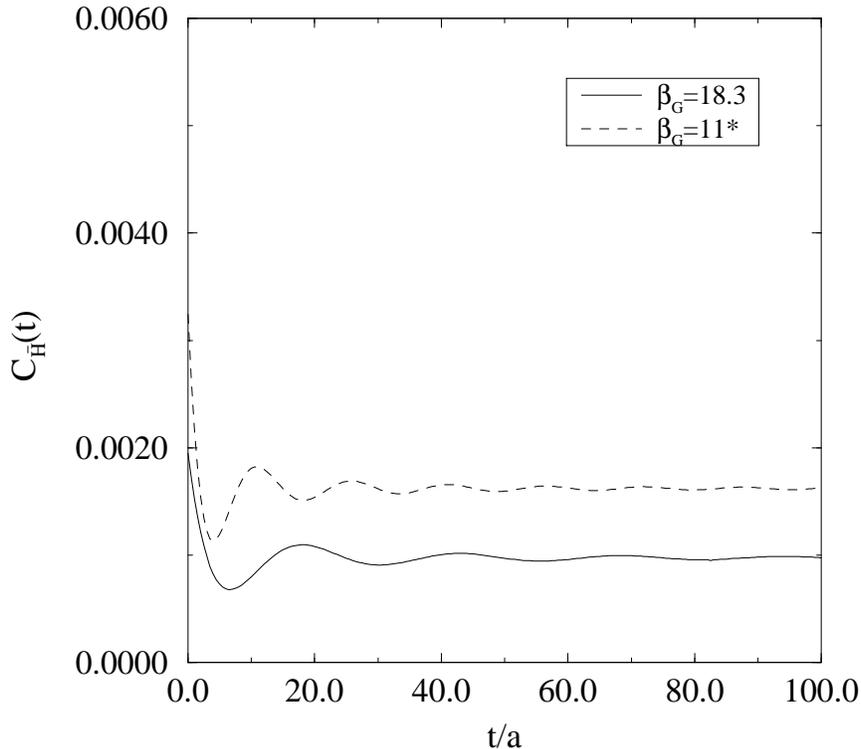}}
 
\vspace*{-6cm}
 
\caption[a]{
The leading order 
perturbative correlator $C_{\bar H}(t)$ in lattice units 
in the symmetric phase at $T/T_c=1.52$. To be compared
with Fig.~8 in~\cite{tang}. The oscillation frequency 
is $2\omega_H$ (this agrees at leading order with the
static screening mass).}
\la{ch}
\end{figure}

\begin{figure}[tb]
 
\vspace*{-1.0cm}
 
\hspace{1cm}
\epsfysize=18cm
\centerline{\epsffile{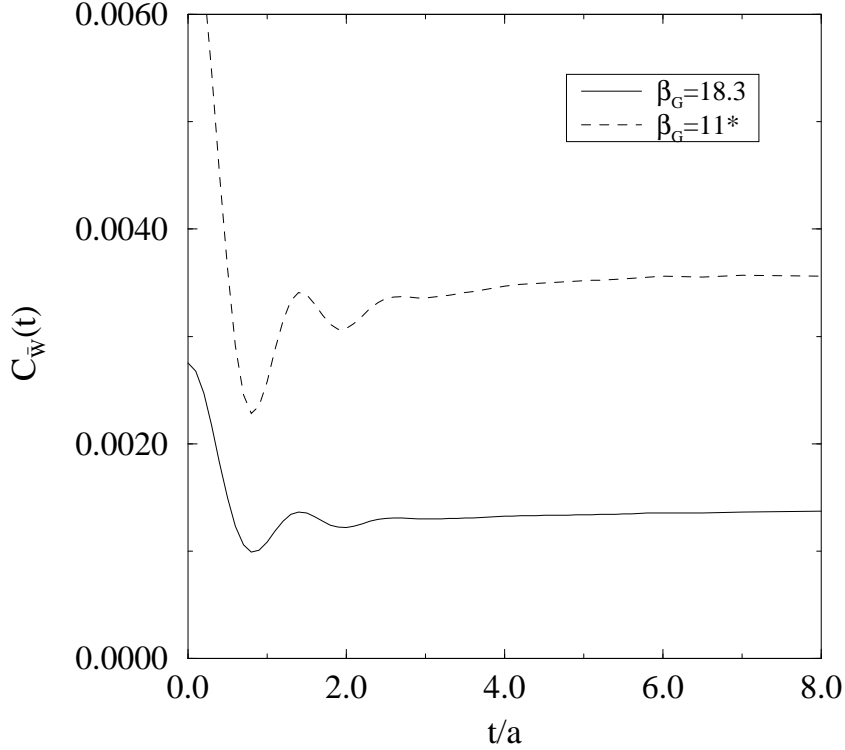}}
 
\vspace*{-6cm}
 
\caption[a]{
The leading order 
perturbative correlator $C_{\bar W}(t)$ 
in lattice units 
in the symmetric phase at $T/T_c=1.52$. To be compared
with Fig.~10 in~\cite{tang}. The oscillation period here
is proportional to the lattice spacing $a$.}
\la{cw}
\end{figure}

The correlation function $C_{\bar H}(t)$ is shown in Fig.~\ref{ch}
and $C_{\bar W}(t)$ in Fig.~\ref{cw}. These should be compared with
Figs.~8, 10 in~\cite{tang}, respectively. It is seen that the
qualitative features can be 
understood quite well with the leading order results. 
Note, in particular, that
the scalar correlation function $C_{\bar H}(t)$
in Fig.~\nr{ch} is oscillating with an
amplitude which decreases with time. It should be emphasized that
this decrease is not related to damping (remember that Fig.~\ref{ch}
shows the tree level results without any interactions and damping
occurs only through interactions). The decrease  
is rather due to the fact that
there is a continuous spectrum of frequencies $\omega>2\omega_H$ 
causing a destructive
interference in the phase space integral, eq.~\nr{symch}.  This shows that
it is difficult to determine a damping rate from the gauge invariant
operator $C_{\bar H}(t)$.

Let us then try to estimate how higher order corrections could modify
the qualitative behavior of 
$C_H(t)$.  Consider the effect of self-energy insertions in the scalar
propagators ${\cal D}(k_0,\vec{k})$. 
The self-energy has an
imaginary part.  Therefore the scalar propagator does not have poles
at $k_0=\pm\sqrt{\vec{k}^2+\omega_H^2 }$ (the lowest order result for
$C_H(t)$ is due to these poles). Since the imaginary part of the 
self-energy is small compared with $\sqrt{\vec{k}^2 + \omega_H^2 }$, $C_H(t)$
will nevertheless still be dominated by the region $k_0 \approx
\pm\sqrt{\vec{k}^2+\omega_H^2 }$.  One can therefore approximate the
scalar propagator as
\begin{eqnarray}
        {\cal D} (k_0 + i\epsilon, \vec{k})
        \approx \frac{1}{-(k_0 + i\Gamma_k)^2
        + \vec{k}^2 + \omega_H^2 },
\label{breit}
\end{eqnarray}
where the width $\Gamma_k$ is given by (see, e.g., \cite{jeon})
$\Gamma_k \equiv -{\rm Im}\Sigma(\sqrt{\vec{k}^2 + \omega_H^2
  },\vec{k})/(2\sqrt{\vec{k}^2 + \omega_H^2 })$.  Inserting this into
eq.~(\ref{imDphi2}) and using eq.~(\ref{relation}), we find
\begin{eqnarray}
        C_H(t) =  T^2 \int\frac{d^3k}{(2\pi)^3} 
        \frac{e^{-2\Gamma_k t}}
        { [\vec{k}^2 + \omega_H^2 ][\vec{k}^2 + \omega_H^2 + \Gamma_k^2]}
        \left[1 + \cos\left(2\sqrt{\vec{k}^2 + \omega_H^2} \,t 
        - 2 \alpha_k\right)\right], \la{gCHt}
\end{eqnarray}
where
$\alpha_k = 
\arctan\left({\Gamma_k}/{\sqrt{\vec{k}^2 + \omega_H^2}}
\right)$. 
Thus the effect of damping should show up in Fig.~\ref{ch}
such that the constant part (corresponding to the first term 
in the square brackets in eq.~\nr{gCHt})
decays away at large times.
This is indeed the qualitative behavior observed in~\cite{tang}.

Finally, we consider the transverse gauge field correlator
$\CA(t,{\bf p})$ measured in~\cite{ambjorn}. Gauge fields
can only be defined in a particular gauge, which in~\cite{ambjorn}
was chosen to be the Coulomb gauge.  We let the external momentum
point into the $x_3$-direction, ${\bf p}={\bf e}_3 (2\pi k/L)$, where
$L$ is the spatial extent of the lattice and $k$ is an integer.  Then
$\CA(t,{\bf p})$ is given, e.g., by the correlator of
$A_1^a$,
\begin{eqnarray}
   \CA(t,{\bf p}) = \frac12 \int d^3 x e^{-i\vec{p}\vec{x}}
   \langle A_1^a(t,{\bf x}) A_1^a(0,{\bf 0}) + A_1^a(0,{\bf 0})
    A_1^a(t,{\bf x})
    \rangle. \la{AA}
\end{eqnarray}

Let us first recall some features of $\CA(t,\vec{p})$ and of the
corresponding analytic Green's function $\DA(\omega,\vec{p})$ for
$|\vec{p}|\sim g^2 T$ in the quantum theory. After the HTL
resummation, $\DA(\omega,\vec{p})$ has poles at
$\omega\approx\pm\omega_W$, where $\omega_W$ is given by
eq.~\nr{omegaWT}.  These poles lead to an oscillation of
$\CA(t,\vec{p})$ on the time scale $\sqrt{\hbar}/(gT)$. In addition,
$\DA(\omega,\vec{p})$ has a discontinuity on the real $\omega$-axis
for $\omega < |{\bf p}|$.  This discontinuity is related to Landau
damping and it gives a contribution $f_{\rm Landau}(t,\vec{p})$ to
$\CA(t,\vec{p})$. The function $f_{\rm Landau}(t,\vec{p})$ does not
involve any oscillations and just constitutes a decaying background
for the superimposed plasmon oscillations. The time scale on which
$f_{\rm Landau}(t,\vec{p})$ varies is $\gsim 1/(g^2T)$. 

Higher order
corrections to $\DA(\omega,\vec{p})$, 
which are not included in HTL effective action but are generated
radiatively within that theory, lead to
a damping of the plasmon oscillations. 
The plasmon damping rate $\Gamma$ 
is of order $g^2T$ and 
has been
computed for $\vec{p} = 0$
in ref.~\cite{braaten}. 

These two different damping effects manifest themselves in
the correlator $\CA(t,{\bf p})$ in quite different ways, so that its
functional form is expected to be
\begin{eqnarray}
  \CA(t,{\bf p}) \sim  
  A \exp({-\Gamma t})\cos (\omega_{W} t+\delta) +  f_{\rm Landau}(t,\vec{p}).
        \label{form}
\end{eqnarray}
The time dependence of $f_{\rm Landau}(t,\vec{p})$ becomes
non-perturbative for $t \gsim 1/(\hbar g^4T)$ \cite{arnold96}, where
$f_{\rm Landau}(t,\vec{p})$ is expected to vanish.  This means that
$\CA(t,{\bf p})$ can be computed perturbatively up to a
non-perturbative constant $\CA(0,{\bf p})$ as long as $t\ll 1/(\hbar
g^4T)$.

In the classical lattice gauge theory one expects a similar
qualitative behavior. In the order of magnitude estimates for
the quantum theory one has to replace $\hbar\to Ta$. The analytic
structure of $\DA(\omega,\vec{p})$ is more complicated than in the
quantum case. The HTL resummed $\DA(\omega,\vec{p})$ depends not only
on the magnitude but also on the direction of $\vec{p}$
\cite{bodeker}.  In particular, there are
directions of $\vec{p}$ for which there is a discontinuity for
arbitrarily large values of $\omega$ \cite{arnold97}.  Nevertheless, the 
qualitative behavior of $\CA(t,\vec{p})$ should be given by 
eq.~\nr{form}. 

Unfortunately, the numerical results for $\CA(t,\vec{p})$ 
in~\cite{ambjorn} have been normalized to $\CA(0,\vec{p})$ which 
cannot be computed in perturbation theory. Comparing our perturbative
estimates with the non-perturbative results therefore requires some
model assumptions about  $\CA(0,\vec{p})$.

In order to account for the plasmon damping effects, we include the
leading order damping rate $\Gamma$ in the HTL resummed propagator
${\cal D}^{ab}_{ij}(\omega,{\bf p})$.  Since the momentum dependence
of $\Gamma$ is not known we use its value for $\vec{p} = 0$ which is
$\Gamma = 0.176 g^2T$ \cite{braaten}.  We can then write for the
transverse components in eq.~\nr{AA}, analogously to eq.~\nr{breit},
\begin{eqnarray}
   {\cal D}^{ab}_{11}(\omega+i\epsilon,{\bf p}) =
   \frac{\delta^{ab}}{ 
   -(\omega+i \Gamma)^2 + \vec{p}^2 + 
   \Pi_{\rm HTL}^{11}(\omega, {\bf p})}. \la{breit2}
\end{eqnarray}

Consider first the zero external momentum case, ${\bf p}={\bf 0}$. 
Eq.~\nr{breit2}
only makes sense when $\Gamma^2 \ll \vec{p}^2 + 
\mathop{\rm Re} \Pi^{11}_{\rm HTL}(\omega,\vec{p})$.
In the latter term in eq.~\nr{Ctp} there should be no problem, 
since for $\vec{p}=0$,
$\Pi^{11}_{\rm HTL}(\omega,\vec{p}) \to \omega_W^2$, but for
the first term 
(i.e., for the static limit $t=0$) the inequality 
is not satisfied
(remember that $\Pi_{\rm HTL}^{11}(0, {\bf p})=0$).
One might try to regulate the real part of the gauge fixed self-energy
phenomenologically with a ``magnetic mass'';
letting $M=\sqrt{m_{\rm magn}^2+\Gamma^2+\omega_W^2}$,
$\widetilde M=\sqrt{m_{\rm magn}^2+\omega_W^2}$, it follows
from eq.~\nr{Ctp} that this would give 
\begin{eqnarray}
\CA(0) & = & \frac{T}{m_{\rm magn}^2+\Gamma^2}, \\
\CA(t)- \CA(0) & = & 
-\frac{T}{M^2}+\frac{T}{M\widetilde M}
e^{-\Gamma t}\cos\!
\left(\widetilde M t-
\arctan\frac{\Gamma}{\widetilde M}
\right).
\end{eqnarray}
Parameterized this way, one can indeed find quite reasonable agreement
with Fig.~6 in~\cite{ambjorn}, but only if $\Gamma$ is chosen to have
a large value, $\Gamma \sim (0.8\ldots 1.0) g^2T$.  Otherwise one is
getting too many oscillations in $\CA(t)$, not observed
in~\cite{ambjorn}. For these large values of $\Gamma$, the magnetic
mass parameter is in fact favored to be small or zero. However, these
fits are quite phenomenological, and thus we will not consider them
any more.

For ${\bf p}\neq 0$, the $t=0$ --part of the leading order
correlator is still parametrically non-perturbative, but at least it is
formally finite for $\Gamma\to 0$ so that 
the perturbative approximation 
might be numerically reasonable.
To get a feeling about the momentum scales in question, note that
for the value $\beta_G=14$ considered in~\cite{ambjorn}, 
$|\vec{p}|=0.69g^2T\times k$ and 
$\omega_W=0.87g^2T$ (the latter can be obtained from eq.~\nr{wbW}
with $m_W(T)=0$).
We have evaluated numerically both the HTL self-energy in 
eq.~\nr{PiW} and 
the remaining $\omega_1$-integral in eq.~\nr{Ctp}, 
for $k=1,\ldots,4$.
The resulting correlators are shown in Fig.~\ref{Acorr}.
This should be compared with Fig.~6 in~\cite{ambjorn}.

\begin{figure}[tb]
 
\vspace*{-1.0cm}
 
\hspace{1cm}
\epsfysize=18cm
\centerline{\epsffile{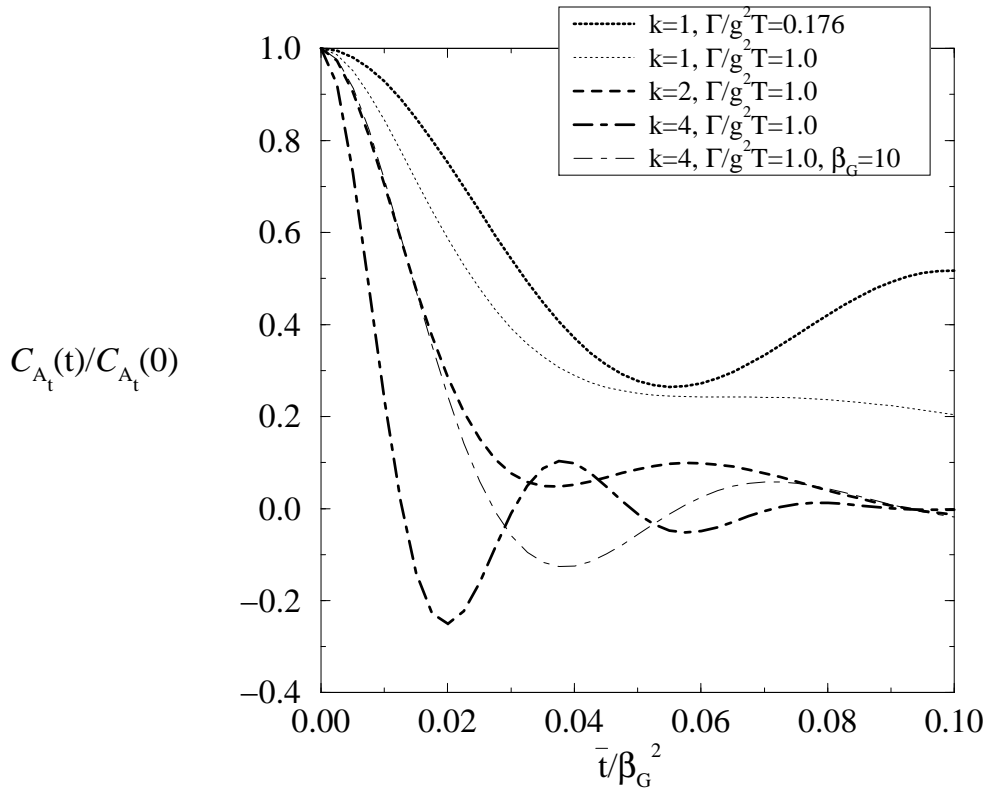}}
 
\vspace*{-6cm}
 
\caption[a]{
The gauge field correlator in the pure SU(2) theory. 
To be compared with Fig.~6 in~\cite{ambjorn}. Here $t$
is expressed in lattice units: $\bar t\equiv t/a$.
Unless otherwise stated, $\beta_G=14$.} 
\la{Acorr}
\end{figure}

The main effects to be seen 
in Fig.~\ref{Acorr} are the following. 
For $k=1$ one can see an oscillation in Fig.~\ref{Acorr}, 
in contrast to Fig.~6 in~\cite{ambjorn}. Thus 
one could say that the non-perturbative
plasmon damping rate is larger than the perturbative estimate
$\Gamma = 0.176 g^2T$. Indeed, 
one has to go to a much larger value, $\Gamma\sim 1.0 g^2T$
($>|\vec{p}|, \omega_W$), 
to get enough plasmon damping. Another observation to be made
at $k=1$ is that in~\cite{ambjorn} the correlator is already
very close to zero at $\bar t/\beta_G^2=0.10$. This is not quite so in 
Fig.~\ref{Acorr}, but one has to wait much longer for the
correlator to vanish. Thus the non-perturbative Landau damping
effects seem also to be larger than the leading order perturbative 
HTL result. At $k=2$ there is no large difference in
Landau damping any more, but the plasmon damping appears
to be somewhat too weak even with $\Gamma\sim 1.0g^2T$.
Finally, at $k=4$ one gets reasonable
agreement between the perturbative and lattice results, 
provided that $\Gamma\sim 1.0g^2T$. One can also 
see that the $\beta_G$-dependence is reproduced; 
the plasmon frequency thus diverges according to eq.~\nr{omegaWa}.

We have also tried the expression $-\omega^2-2i\omega\Gamma$
in the denominator of eq.~\nr{breit2}, so that the real part
is not modified by $\Gamma^2$. The qualitative conclusions 
and the preferred value $\Gamma\sim 1.0g^2T$ remain the same. 
Based on those curves, one would nevertheless say that even 
at $k=2$ there is too little Landau damping in the 
perturbative estimates, but at $k=4$ one again gets good
agreement.

\subsubsection*{Summary and Conclusions}

We have computed several quantities related to real time correlation
functions in the classical SU(2) and SU(2)+Higgs models on the lattice,
using hard thermal loop resummed perturbation theory.

Our results for the gauge field and scalar plasmon frequencies in the
broken phase are in remarkable agreement with the numerical lattice
simulations in ref.~\cite{tang}. We have reiterated that the classical
gauge field plasmon frequency is divergent in the continuum limit and
we have demonstrated that this is consistent with the results of
ref.~\cite{tang}, where 
the plasmon frequency was claimed to be lattice spacing
independent.

For the symmetric phase we have computed gauge invariant scalar
and vector correlators as functions of time at the lowest order
in perturbation theory. Furthermore, we have estimated the effect of
higher order corrections.  Our results are in good 
agreement with ref.~\cite{tang}. 
We have shown that it is difficult to extract damping rates from 
the measurement of these correlators.

Finally, we have studied the correlator of the transverse gauge field
in pure SU(2) gauge theory. While the qualitative features of the
numerical simulations in ref.~\cite{ambjorn} are consistent with our
perturbative estimates, there appear to be significant quantitative
discrepancies. The damping of the plasmon oscillations observed
in~\cite{ambjorn} appears to be much stronger than one would expect
from the perturbative result for the damping rate~\cite{braaten}.
This is puzzling because the damping rate is of the order $g^2 T$ and
should therefore have a classical continuum limit.  However, 
one should keep in mind
that the perturbative estimates are reliable only if the
plasmon frequency is much larger than the damping rate which is not
the case for the lattice spacing used in ref.~\cite{ambjorn}.

{\bf Acknowledgments.} 
We are grateful to E.~Berger, P.~Overmann, O.~Philipsen, 
M.G.~Schmidt and
I.O.~Stamatescu for useful discussions.

\subsubsection*{Note added}

After this paper was submitted,
we were informed by J.Smit that 
in the revised version of Ref.~\cite{tang}, 
Tang and Smit have weakened
their conclusions concerning the lattice spacing independence of
the $W$ plasmon frequency in the broken phase, so that their
statements are now in better accordance with ours. We thank 
J.Smit for communication on this issue. 

In view of the interpretation in~\cite{ay2}, 
let us stress that the discussion around eq.~\nr{breit}
is not meant to be a consistent quantitative 
estimate of the higher order corrections. We just want to
see in which qualitative way the higher order corrections
might manifest themselves.

\end{document}